\RequirePackage{fixltx2e}
\documentclass[aps,prl,reprint,superscriptaddress,floatfix]{revtex4-1}
\usepackage{graphicx,graphics,amsmath,amssymb,rotate,textcomp,gensymb}
\usepackage{mathrsfs,float}
\usepackage[T1]{fontenc}
\usepackage{dcolumn}
\usepackage{wrapfig}
\usepackage{caption,subcaption}
\usepackage[none]{hyphenat}
\usepackage{hyperref}
\usepackage{times}
\usepackage{natbib}

\begin{document}

\title{Novel chiral smectic phase generation in systems of polar ellipsoidal molecules with reflection asymmetry: A molecular dynamics simulation study }

\author{Tanay Paul}
\email{tanaypaul9492@gmail.com}
\author{Jayashree Saha}
\email{jsphy@gmail.in}
\affiliation{Department of Physics, University of Calcutta,92, A. P. C. Road, Kolkata - 700009, India.}

\date{\today}


\begin{abstract}
Computer simulation study of phase transitional behaviour of cholesterol molecules embedded with terminal dipole is reported. In this work, coarse-grained modeling of cholesteric molecules is done to study the influence of the coupled chiral and dipolar interactions on macroscopic liquid-crystalline phase formation. This NVT molecular dynamics simulation study demonstrates the formaton of novel Smectic Blue phase (BP\textsubscript{Sm}) which is a recent experimentally discovered chiral phase. Our study reveals that the higher strength of chiral interaction induces blue phase, whereas, larger dipolar interaction can bring bilayered smectic blue phase.
\end{abstract}

\maketitle

\section{Introduction}
Chirality, a property of reflection asymmetry pervades many branches of science, ranging from elementary particle physics to biological sciences. The word \textquoteleft Chirality\textquoteright, first introduced by Lord Kelvin (1894), has shown important manifestations in microscopic as well as in macroscopic phenomena. Molecular homochirality, i.e. presence of only single handedness, is absolutely essential for any living system. A molecule is called chiral if it is not superimposable into its mirror image. Its symmetry group does not contain the element S\textsubscript{n} - rotation around an S\textsubscript{n} axis by $\frac{2\pi}{n}$ followed by a mirror reflection through a plane perpendicular to that axis. A chiral moleule is three dimensional and not uniaxial. Molecular chirality induces chiral interaction that gives rise to liquid crystalline cholesteric phases.

There are many successful computer simulation (using both coarse-grained and all-atom model) studies on different liquid crystal phases e.g. nematic and smectic phases showing variety of different phenomena like ferroelectricity, bilayer phase formation etc., but so far there are not sufficient theoretical study on chiral liquid crystal phases. There are many experimental studies revealing the existence of several novel phases generated by chiral molecules like cholesteric phases without positional order (N\textsuperscript{\textasteriskcentered}) , blue phases (BP I, BP II, BP III), helical smectic A\textsuperscript{\textasteriskcentered} phase, tilted smectic C\textsuperscript{\textasteriskcentered} phase, twisted grain boundary phases (TGB) etc. Computer simulation studies based on proper modelling of the chiral interaction is necessary to assess basic interactions responsible for the generation of these phases and to develop understanding the link between their microscopic and macroscopic properties. Though there are a few theoretical analyses of classical \cite{straley, evans, harris} and molecular-statistical \cite{goosens, van-der-meer, shroder} origin of chiral interactions, computer simulation studies which focuss on essential realistic interactions are sparse. A simulation study of cholesteric phase incorporating lattice model considering all rotational degrees of freedom was reported by Saha et al \cite{saha} . Memmer et al. presented a series of Monte-Carlo simulation results of chiral liquid crystal phases including blue phase \cite{memmer93, memmer2k} considering a coarse-grained model representing chiral Gay-Berne fluid with an additive chiral interaction term.

Chirality-induced celebrated macroscopic liquid crystal phases e.g. helical nematic phase and optically isotropic blue phases are formed depending on the amount of the chirality in the system. Generally the blue phases are obtained with sufficiently large chirality. While BP I has a body centered cubic symmetry and BP II has a simple cubic symmetry, structure of BP III is relatively less known to the extent that it is amorphous having short distance order. Double twist cylinders form periodic cubic lattice of BP \cite{crooker, dabrowski, dierking}. In a double twist cylinder structure the lowest energy local director rotates about any radius of the cylinder and is parallel to the axis of the cylinder at the center. Thus this type of arrangement inevitably forms defects in the orientation of the molecules, to reduce elastic strain energy, as fitting such double twist cylinders in three dimensional space is impossible maintaining the matching of local directors everywhere. A balance between the defects and lowest energy local directors induces the formation of BPs. Blue phases are thermodynamically different from each other and first order phase transitions occur between these phases.

A recently discovered \cite{grelet} mesophase of thermotropic chiral liquid crystals is smectic blue phase (BP\textsubscript{Sm}) which is the result of two competing frustrated mesophases, one is twist-grain-boundary (TGB) phase which is the consequence of the balance between helical twist and the smactic order, and the other is the classical blue phase. Goodby et al \cite{goodby} suggested that the TGB model, a phase predicted by Renn and Lubensky \cite{renn}, is suitable for Smectic A\textsuperscript{\textasteriskcentered}, where the grain boundaries of periodic arrangements of screw dislocations seperate different blocks consisted of Smectic A layers. Smectic blue phase exhibits double frustration - that of blue phase with twist at the molecular level, where frustration occurs due to the impossibility of fitting molecular chirality in three dimension and that of TGB phase where helical twist occurs at a macroscopic length scale of smectic slabs.

Although different structures of liquid crystalline phases of rod-like and disc-like chiral mesogens are well observed and characterized, the link between chirality in microscopic and macroscopic level is little understood \cite{lubensky}. For the binary mixtures of enantiomers of a chiral compound a typical phase diagram in the temperature-chirality parameter plane shows the phase transitions from isotropic liquid phase to smectic phase which has positional ordering where blue phase occurs as an intermediate phase and cholesteric phase where the blue phase exists for relatively large temperature range for the greater relative abundance of a single component chiral enantiomer \cite{slaney}. The existence of blue phase, the characterization of its structural properties and dependence of occurence of different blue phase on the strength of chiral interaction on a large system size was investigated by Memmer \cite{memmer2k} .

The cholesterols in mixture with phospholipids can form liquid crystalline bilayered phases \cite{marsh, mateo}. Both phospholipid and cholesterol molecules have polar parts, but the dipole moment of the latter is of relatively smaller magnitude compared with that of the polar head-group of a phospholipid. Dipolar strengths of some cholesteryl compounds are measured by Gopalakrishna et al. \cite{gopalakrishna}. Cholesterol molecules themselves are chiral molecules and form chiral phases. To the best of our knowledge no one studied theoretically the effect of coupled chiral and dipolar interaction on structure of liquid crystal phases. In this work Molecular Dynamics simulation has been performed to investigate the link between molecular chirality jointly with dipolar interactions of cholesterols in microscopic level and macroscopic chiral liquid crystal phases, considering a coarse-grained modelling of cholesterol molecules.

\section{Model and Computational Details}
In our coarse-grained Molecular Dynamics simulation study, cholesteric molecules are modelled as molecules of ellipsoidal shape having a single terminal point dipole. The orientation of the dipole is fixed with respect to the molecular long axis. The total interaction between the two molecules $i$ and $j$, the centers of mass of which are separated by $\vec{r}_{ij}$ with long axes unit vectors $\hat{u}_{i}$ and $\hat{u}_{j}$, is represented by the pair potential,
\begin{eqnarray}
U(\vec{r}_{ij},\hat{u}_{i},\hat{u}_{j})&=& -c\cdot U_{C}(\vec{r}_{ij},\hat{u}_{i},\hat{u}_{j}) + U_{GB}(\vec{r}_{ij},\hat{u}_{i},\hat{u}_{j}) \nonumber \\
& & + U_{dd}(\vec{r}_{d},\hat{u}_{d_{i}},\hat{u}_{d_{j}}) \label{eq:1}\\
&=& -c4\epsilon(\hat{r}_{ij}, \hat{u}_{i}, \hat{u}_{j})\rho_{ij}^{-7}\{(\hat{u}_{i}\times\hat{u}_{j})\cdot\hat{r}_{ij}\}(\hat{u}_{i}\cdot\hat{u}_{j}) \nonumber \\
& & + 4\epsilon(\hat{r}_{ij}, \hat{u}_{i}, \hat{u}_{j})(\rho_{ij}^{-12}-\rho_{ij}^{-6}) \nonumber \\
& & + \frac{1}{r^3_{d}}[\vec{\mu}_{d_{i}}\cdot\vec{\mu}_{d_{j}}-\frac{3}{r^{2}_{d}}(\vec{\mu}_{d_{i}}\cdot\vec{r}_{d})(\vec{\mu}_{d_{j}}\cdot\vec{r}_{d})] \label{eq:2}
\end{eqnarray}
Here $U_C$ is the chiral interaction pair potential, a pseudo-scalar \cite{memmer93} , which produces intermolecular torque of a particular handedness depending on the sign of chirality strength parameter $c$, the value of which is zero for achiral molecules and the achiral part $U_{GB}$ \cite{gb} is a van-der-Waals type interaction for anisotropic molecules represented by Gay and Berne, whereas, the electrostatic interaction between the dipolar portions of the cholesteric molecules is represented in our simulation simply by the dipole-dipole interaction $U_{dd}$ which acts between the two point dipoles. Here, $\vec{r}_{d}=r_{d}\hat{r}_{d}$ is the separation vector joining the two point dipoles embedded on molecules $i$ and $j$, and the dipole moment vectors of the point dipoles embedded on molecules $i$ and $j$ are $\vec{\mu}_{d_{i}}\equiv \mu^*\hat{u}_{d_{i}}$ and $\vec{\mu}_{d_{j}}\equiv \mu^*\hat{u}_{d_{j}}$  respectively where $\mu^*= (\mu^2/\varepsilon_{s}\sigma_{0}^3)^{1/2}$ is the dimensionless dipole moment of the point dipoles. The point dipoles are fixed on the long axes of each of the molecules at a distance of $\sigma_0$ from the center of mass of the molecules and the dipole moment vectors are perpendicular to the long axes of the molecules. Reaction Field method \cite{onsager} is considered to incorporate the long-range nature of the dipolar interaction.

The form of the chiral interaction potential can be obtained from the multipole expansion of electrostatic interaction \cite{van-der-meer}, while the separation and orientation dependent factor $\epsilon(\hat{r}_{ij}, \hat{u}_{i}, \hat{u}_{j})\rho_{ij}^{-7}$ is chosen as of Gay-Berne type, the term $\{(\hat{u}_{i}\times\hat{u}_{j})\cdot\hat{r}_{ij}\}(\hat{u}_{i}\cdot\hat{u}_{j})$ induces twist angle between molecules and is responsible for generating cholesteric phase.
For two molecules $i$ and $j$ placed side-by-side i.e. $\hat{r}_{ij}$ is $\perp$ to both $\hat{u}_i$ and $\hat{u}_j$ the term $\{(\hat{u}_{i}\times\hat{u}_{j})\cdot\hat{r}_{ij}\}(\hat{u}_{i}\cdot\hat{u}_{j})$ is minimized at an twist angle of $\pi/4$ between the long axes of the two molecules, thus induces a twist between molecules arranged side-by-side. But, for two molecules arranged end-to-end, $\hat{r}_{ij}$ and $(\hat{u}_{i}\times\hat{u}_{j})$ are $\perp$ to each other and hence the term $\{(\hat{u}_{i}\times\hat{u}_{j})\cdot\hat{r}_{ij}\}(\hat{u}_{i}\cdot\hat{u}_{j})$ becomes zero, making the chiral potential part zero and then only the achiral part alongwith dipolar interaction is applied between two such molecules favouring nematic arrangement.

The separation dependent term in both $U_C$ and $U_{GB}$ contains $\rho_{ij} = [ r_{ij} - \sigma(\hat{ r}_{ij}, \hat{u}_{i}, \hat{u}_{j}) + \sigma_{0}] / \sigma_{0}$. Here the orientation dependent range parameter $\sigma$ is given by,
\begin{equation}
\begin{aligned}
{}&\sigma\left(\hat{ r}_{ij},\hat{u}_{i},\hat{u}_{j}\right)=\sigma_{0}\times \\
  & \left \{1-\frac{\chi}{2}\left[\frac{(\hat{u}_{i}\cdot\hat{r}_{ij}+\hat{u}_{j}\cdot\hat{r}_{ij})^2}{1+\chi(\hat{u}_{i}\cdot\hat{u}_{j})}+\frac{(\hat{u}_{i}\cdot\hat{r}_{ij}-\hat{u}_{j}\cdot\hat{r}_{ij})^2}{1-\chi(\hat{u}_{i}\cdot\hat{u}_{j})}\right]\right \}^{-{\frac{1}{2}}} \label{eq:3}
\end{aligned}
\end{equation}
where $\chi$ is related to the shape anisotropy, $\kappa \equiv \left(\frac{\sigma_{e}}{\sigma_{0}}\right)$ of the particles,
\begin{equation}
\chi=\frac{\kappa^2-1}{\kappa^2+1} \label{eq:4}
\end{equation}
Here $\sigma_{e}$, $\sigma_{0}$ are size parameters which are basically the length and the breadth of the particles. The energy term in equation (\ref{eq:2}) can be expressed as,
\begin{equation}
\epsilon(\hat{r}_{ij},\hat{u}_{i},\hat{u}_{j})=\varepsilon_{0}\varepsilon'^{\mu}\left(\hat{r}_{ij},\hat{u}_{i},\hat{u}_{j}\right)\varepsilon^{\nu}\left(\hat{u}_{i},\hat{u}_{j}\right) \label{eq:5}
\end{equation}
where, $\varepsilon_0$ is a strength constant which is taken as $1$ in our study and the orientation dependent strength parameter adjusting the size anisotropy of molecules is
\begin{equation}
\varepsilon\left(\hat{u}_{i},\hat{u}_{j}\right)=\left[1-\chi^2\left(\hat{u}_{i}\cdot\hat{u}_{j}\right)^2\right]^{-{\frac{1}{2}}} \label{eq:6}
\end{equation}
and the strength parameter adjusting the end-to-end and side-by-side well depths with respective orientations of the molecules of the pair is,
\begin{equation}
\begin{aligned}
{}&\varepsilon'\left(\hat{r}_{ij},\hat{u}_{i},\hat{u}_{j}\right)= \\
  & 1-\frac{\chi'}{2}\left[\frac{\left(\hat{u}_{i}\cdot\hat{r}_{ij}+\hat{u}_{j}\cdot\hat{r}_{ij}\right)^2}{1+\chi'\left(\hat{u}_{i}\cdot\hat{u}_{j}\right)}+\frac{\left(\hat{u}_{i}\cdot\hat{r}_{ij}-\hat{u}_{j}\cdot\hat{r}_{ij}\right)^2}{1-\chi'\left(\hat{u}_{i}\cdot\hat{u}_{j}\right)}\right] \label{eq:7}
\end{aligned}
\end{equation}
The parameter $\chi'$ reflects the anisotropy in the attractive forces,
\begin{equation}
\chi'=\frac{1-\kappa'^{\frac{1}{\mu}}}{1+\kappa'^{\frac{1}{\mu}}} \label{eq:8}
\end{equation}
where $\kappa'$ is the well depth anisotropy ratio: $ \kappa' = \frac{\varepsilon_{e}}{\varepsilon_{s}} $ and $\varepsilon_{e}$, $\varepsilon_{s}$ are the well depths for the end-to-end and side-by-side configurations. The parameters $\kappa=3$, $\kappa'=1/5$, $\mu=1$, $\nu=2$ were considered in our study.

NVT molecular dynamics simulation method is used in our study for molecules of size assymmetry ratio $\kappa=3$ with scaled density $\rho^*\left(\rho^*\equiv\frac{N\sigma_{0}^3}{V}\right)$ set to $0.30$. For all systems simulation run has been started from a well equilibrated isotropic phase and then decreasing scaled temperature ( $T^*\equiv k_{B}T/\varepsilon_0$, $k_B$ being the Boltzmann constant ) gradually in order to obtain more ordered phase and at each temperature stage the configuration in previous tempearature stage has been used as the initial configuration. For a fixed chirality strength parameter $c$ the scaled dipole moment $\mu^*$ are varied to study the effect of dipole moment on the phase behaviour of cholesteric molecules. For each set of values of $c$ and $\mu^*$ simulation run is started from a typical isotropic phase and then the scaled temperature $T^*$ is decreased to get more ordered phase and at each temperature stage the final configuration is obtained starting from the configuration obtained at a higher temperature state. At a particular temperature stage to get equilibrium configuration a run of $10^6$ steps is performed and the averages are calculated over last $10^5$ steps.

\section{Results}

\begin{table}[ht]
 \centering
 \begin{tabular}{c|c|ccccccc}
 \hline
 $c$       & $0.5$   &       &       &       & $1.0$ &       &       &       \\
 \hline
 $\mu^*$   & $1.0$   & $0.1$ & $0.3$ & $0.5$ & $0.7$ & $0.9$ & $1.0$ & $1.4$ \\
 \hline
 Phases    & Smectic &BP\textsubscript{Sm}&BP\textsubscript{Sm}&BP\textsubscript{Sm}&BP\textsubscript{Sm}&BP\textsubscript{Sm}&BP\textsubscript{Sm}&BP\textsubscript{Sm} \\
           & A\textsuperscript{*} &       &       &       &       &       &       &  (Bilayered)   \\
 \hline
 $T^*$     &   0.8   &  0.9  &  1.0  &  1.0  &  1.1  &  1.2  &  1.3  &  1.5  \\
 \hline

 \end{tabular}
\caption{Different phases\label{table:1}}
\end{table}

\begin{figure}[htp]
\begin{subfigure}{0.155\textwidth}
 \includegraphics[width=\textwidth]{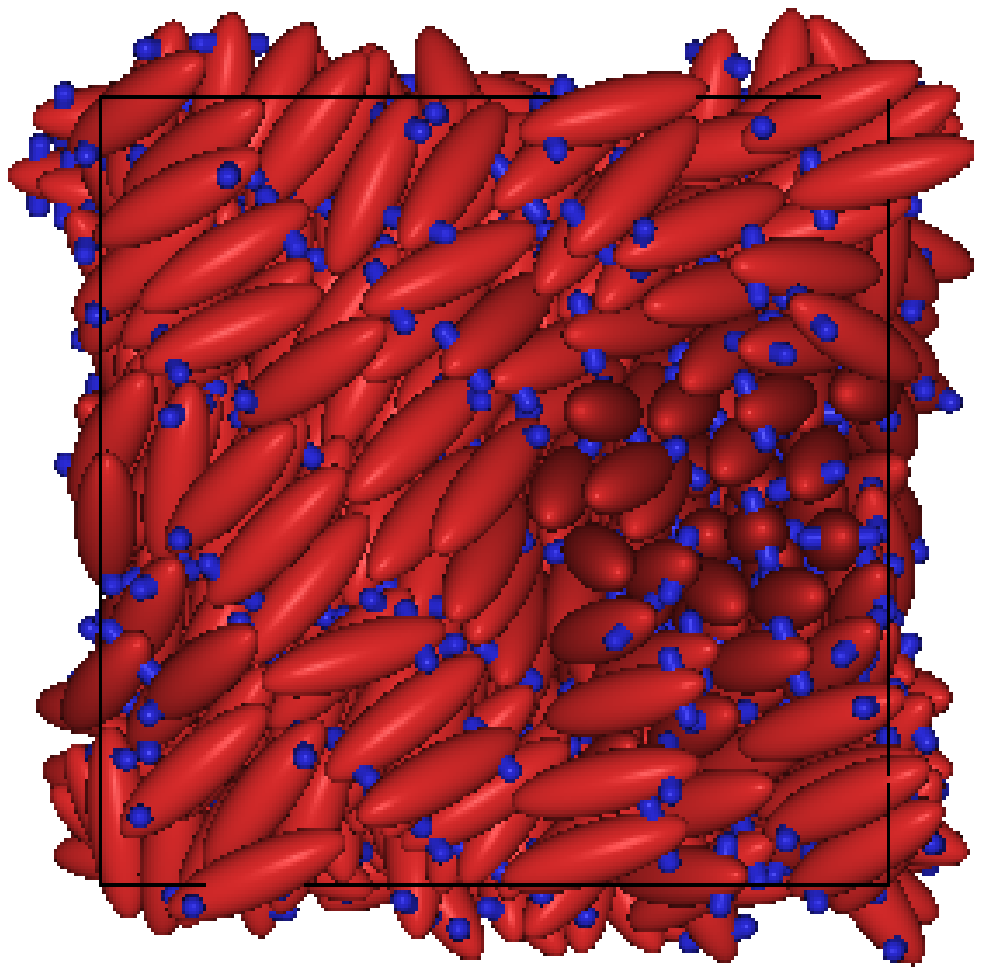}
 \caption{\label{fig:m0.1}}
\end{subfigure}
\begin{subfigure}{0.155\textwidth}
 \includegraphics[width=\textwidth]{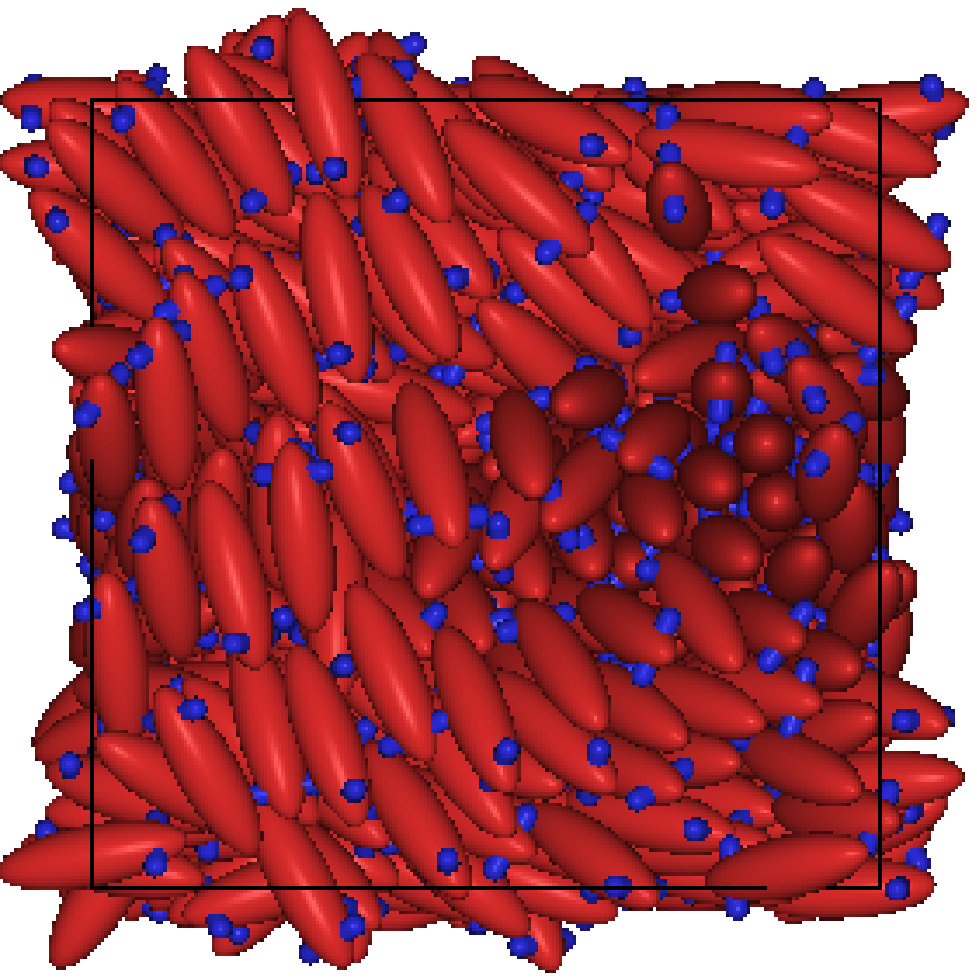}
 \caption{\label{fig:m0.3}} 
\end{subfigure}
\begin{subfigure}{0.155\textwidth}
 \includegraphics[width=\textwidth]{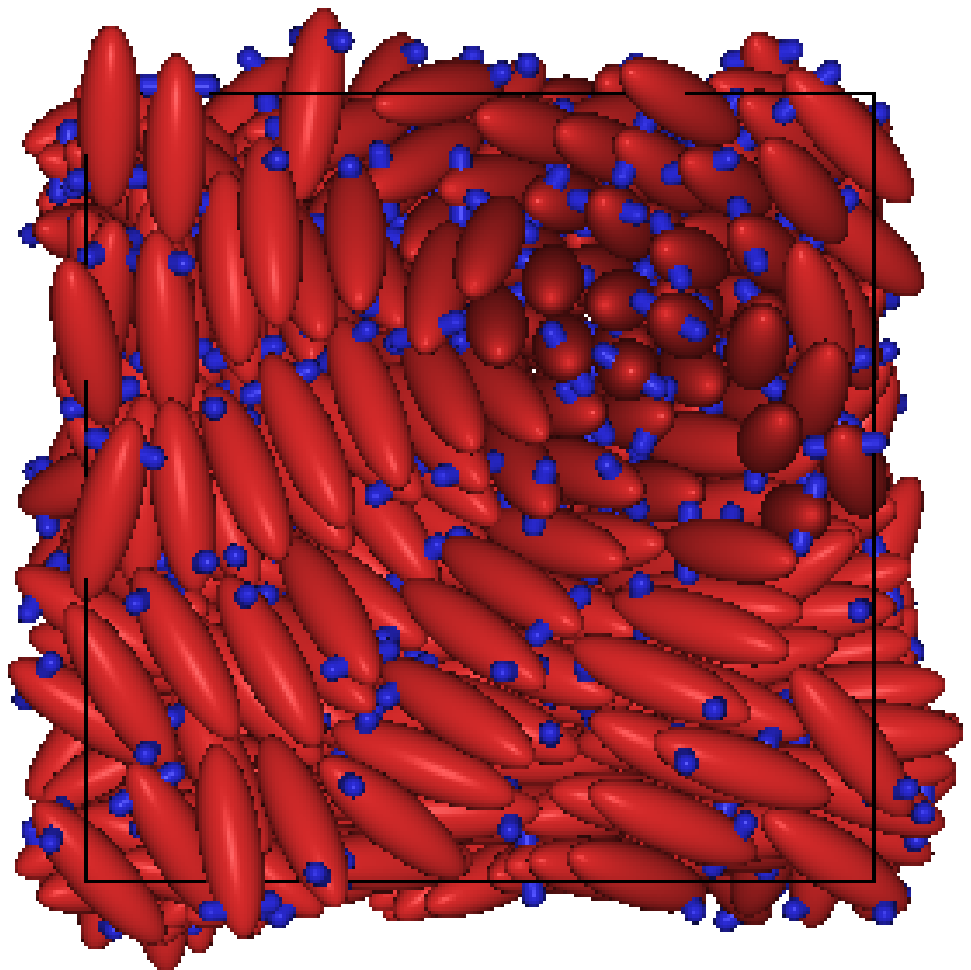}
 \caption{\label{fig:m0.5}}
\end{subfigure}
\\
\begin{subfigure}{0.155\textwidth}
 \includegraphics[width=\textwidth]{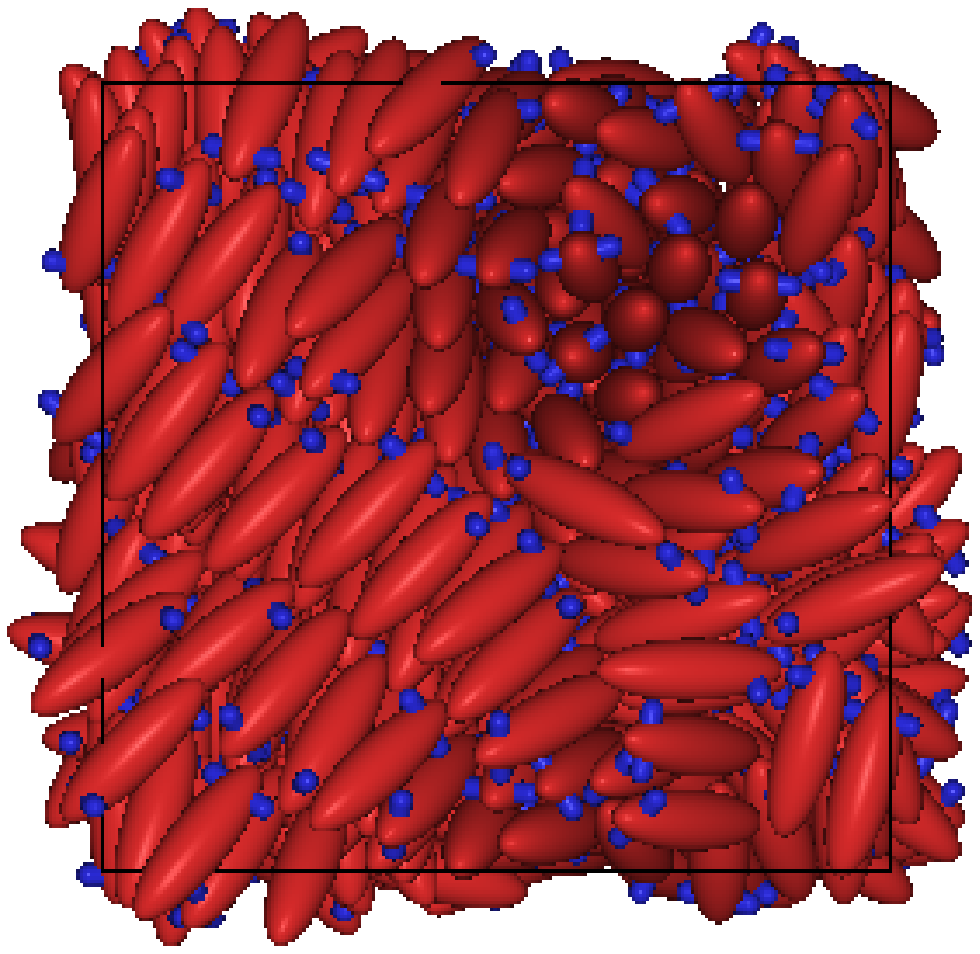}
 \caption{\label{fig:m0.7}}
\end{subfigure}
\begin{subfigure}{0.155\textwidth}
 \includegraphics[width=\textwidth]{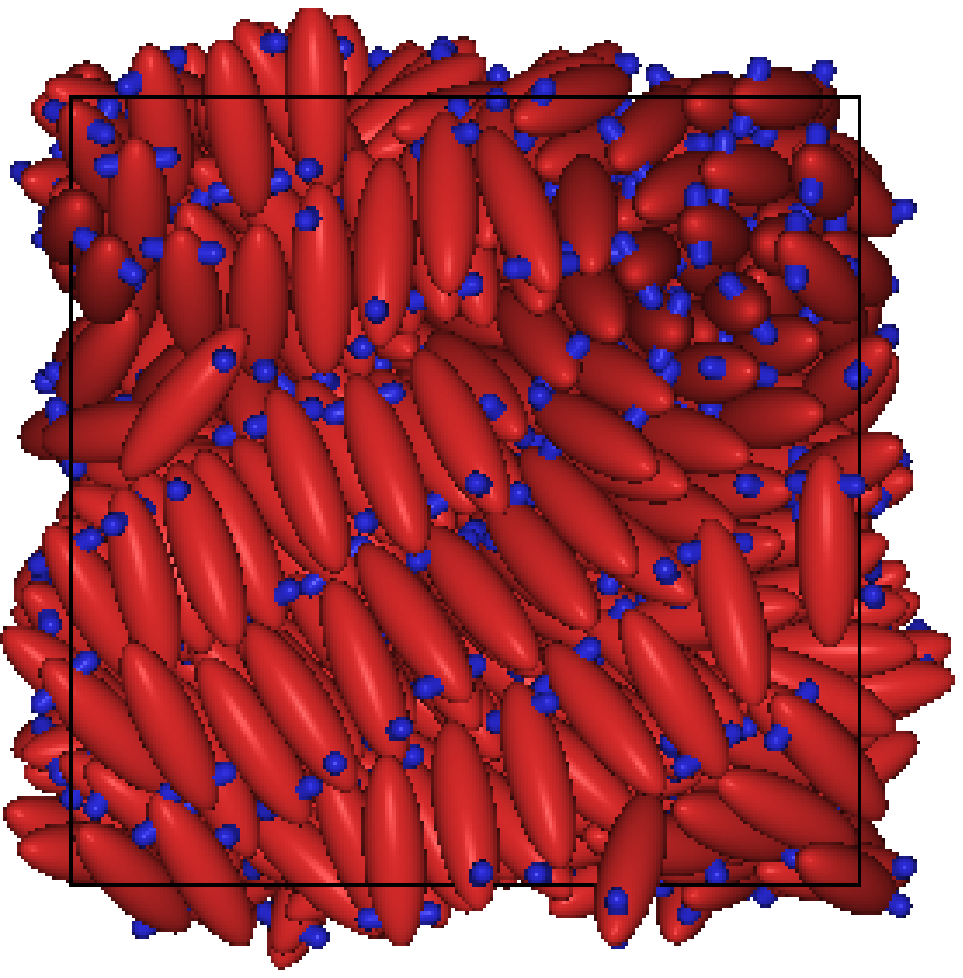}
 \caption{\label{fig:m0.9}}
\end{subfigure}
\begin{subfigure}{0.155\textwidth}
 \includegraphics[width=\textwidth]{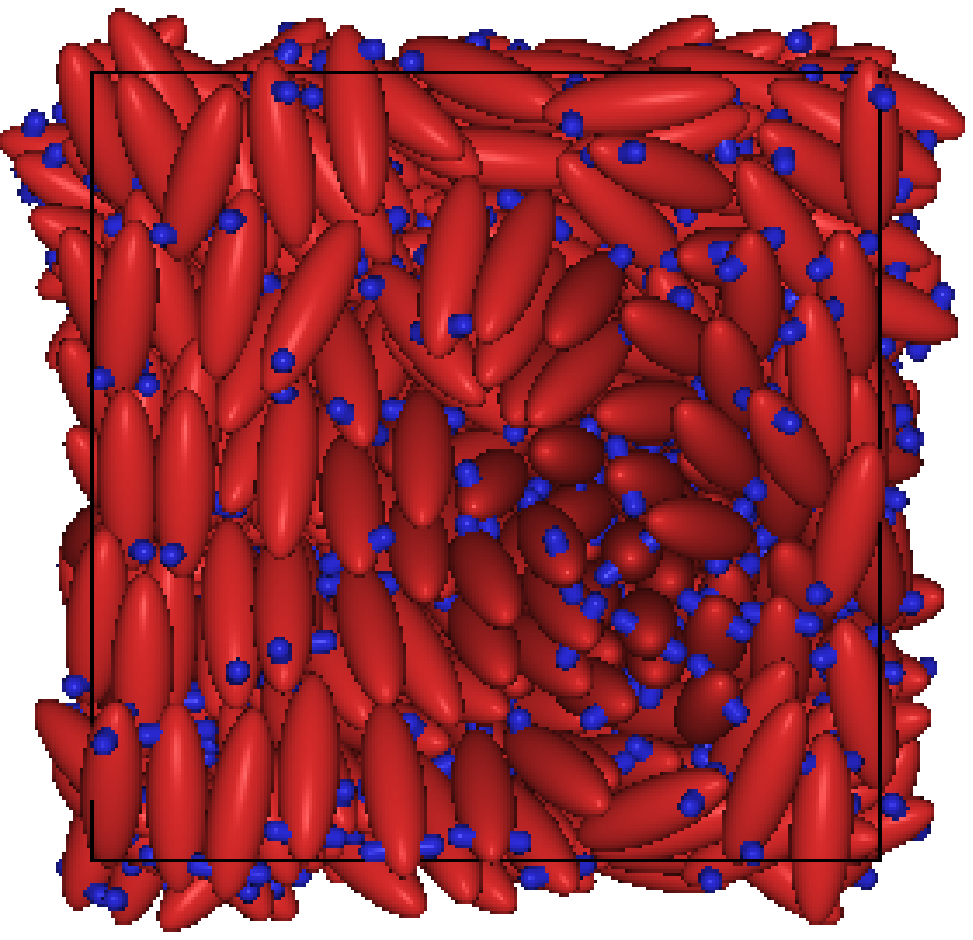}
 \caption{\label{fig:m1.0}}
\end{subfigure}
\\
\begin{subfigure}{0.155\textwidth}
 \includegraphics[width=\textwidth]{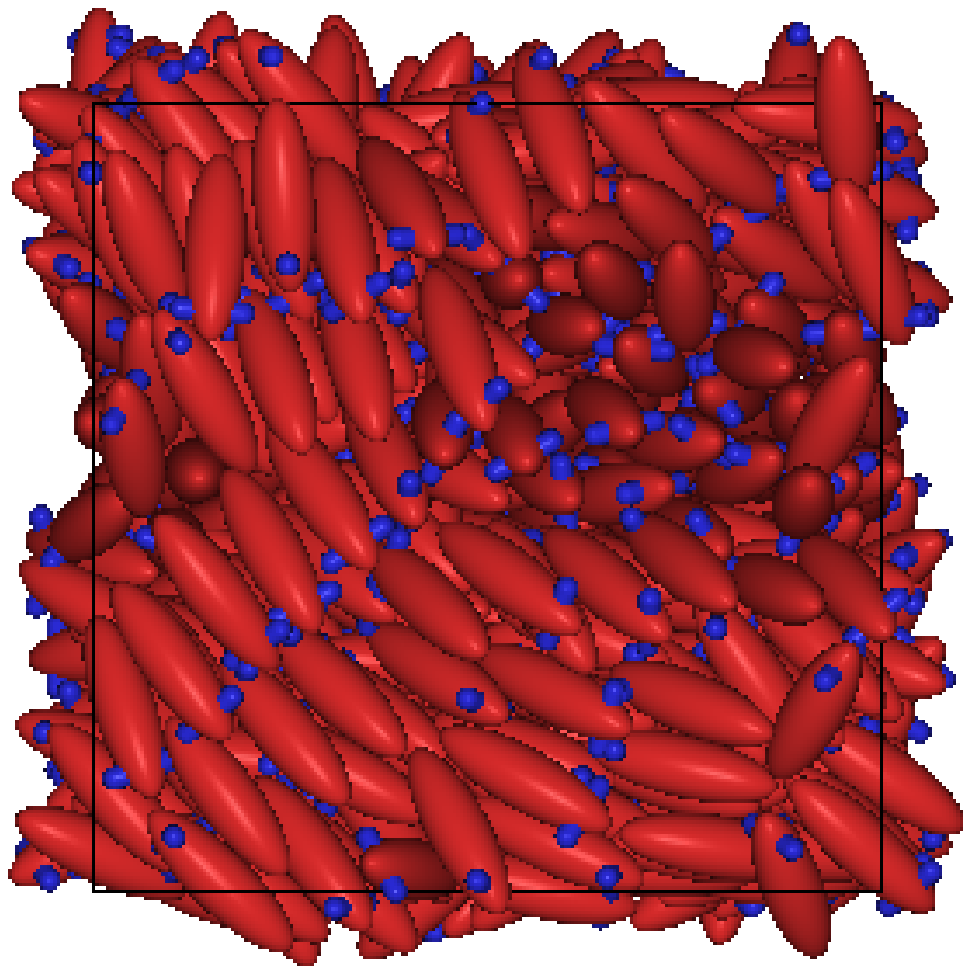}
 \vspace{0.5cm}\caption{\label{fig:m1.4}}
\end{subfigure}
\caption{Visualization of the stable configurations of the blue phases obtained for $N=500$ moleules: (a) $\mu^*=0.1$, (b) $\mu^*=0.3$, (c) $\mu^*=0.5$, (d) $\mu^*=0.7$, (e) $\mu^*=0.9$, (f) $\mu^*=1.0$, (g) $\mu^*=1.4$. (Molecules are presented by ellipsoids and positions of the dipoles by short spherocylinders.)\label{fig:qmga}}
\end{figure}

Smectic blue phases ( BP\textsubscript{Sm} ) are formed in a system where the value of $c$ is set to $1.0$ using number of molecules $N=500$. Variation of dipole moment does not affect it much ( Table \ref{table:1} ) as from the table it is clear that for $c=1.0$ and several values of $\mu^*$ from $0.1$ to $1.4$ BP\textsubscript{Sm} is obtained. A typical phase sequence isotropic-BP-BP\textsubscript{Sm} is observed with decreasing temperature. The values of transition temperature $T^*$ for the respective values of $\mu^*$ at which the BP\textsubscript{Sm} phases occurred is given in the table which is greater for higher values of $\mu^*$. To check the dependence of occurrence of BP\textsubscript{Sm} phase on chirality strength parameter $c$, phase sequences were checked with decreasing temperature for a lower value of $c=0.5$ considering a typical value of $\mu^*=1.0$ where blue phase is not obtained, instead helical nematic phase occurred from the isotropic one. In this case, as the temperature is lowered further Smectic A\textsuperscript{*} phase with positional ordering occured. A novel phase is obtained for $\mu^*=1.4$ and $c=1.0$ where bilayered smectic blue phase ( Fig: \ref{fig:m1.4} ) is found to be energetically most favoured one. Surely the high value of dipole moment induces this bilayer. With these values of $c$ and $\mu^*$ i.e. with $c=1.0$ and $\mu^*=1.4$ simulation run was performed in a system with $N=1372$ and same result has been obtained.

\begin{figure}[htp]
\begin{center}
\includegraphics[width=0.45\textwidth]{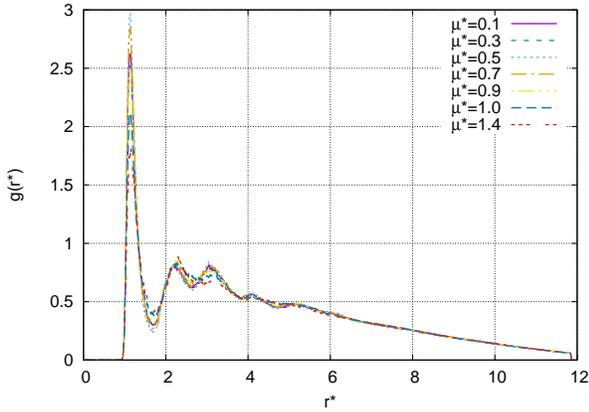}
\caption{Plot of $g(r^*)$ for all the stable configurations with different value of $\mu^*$ showing similar behaviour with short range correlation only.\label{fig:gr}}
\end{center}
\end{figure}

In order to analyse structural properties of the phases obtained in more details some rotationally invariant distribution functions are computed. Radial distribution function or pair distribution function $g(r^*)$ ($r^*$ is the scaled intermolecular seperation) is calculated and plotted in figure \ref{fig:gr} which shows short range positional order but no order in long range. Some selected longitudinal orientational correlation functions $S_{220}(r^*_{\parallel}/d)$ and $S_{221}(r^*_{\parallel}/d)$ are computed to find out orientational correlations as a function of the scaled intermolecular distance $r^*_{\parallel}$ measured along an appropriately chosen reference axis and further scaled by a selected distance $d$ which is related to the periodicity of the phase studied. In the helical nematic phase the helical axis has been chosen as reference axis and the pitch length as the scaling length $d$. In case of blue phases each of three vectors normal to the mutually perpendicular sides of the simulation box are chosen as reference axis and the correlation functions are computed seperately in each cases and twice the sidelength of the simulation box has been chosen as the scaling length $d$. The mathematical form of these functions are given by,
\begin{eqnarray}
S_{220}(r^*_{\parallel}/d)&=&\frac{1}{2\sqrt{5}}\langle 3(\hat{u}_i\cdot\hat{u}_j)^{2}-1\rangle_{ij}\\
S_{221}(r^*_{\parallel}/d)&=&-\sqrt{\frac{3}{10}}\langle[(\hat{u}_{i}\times\hat{u}_{j})\cdot\hat{r}_{ij}](\hat{u}_{i}\cdot\hat{u}_{j})\rangle_{ij}
\end{eqnarray}
where the $ij$ subscript indicates the average over all molecular pairs seperated by a distance $r^*_{\parallel}/d$ along appropriately chosen reference axis.
\begin{figure}[htp]
\begin{center}
 \begin{subfigure}{0.45\textwidth}
 \includegraphics[width=\textwidth]{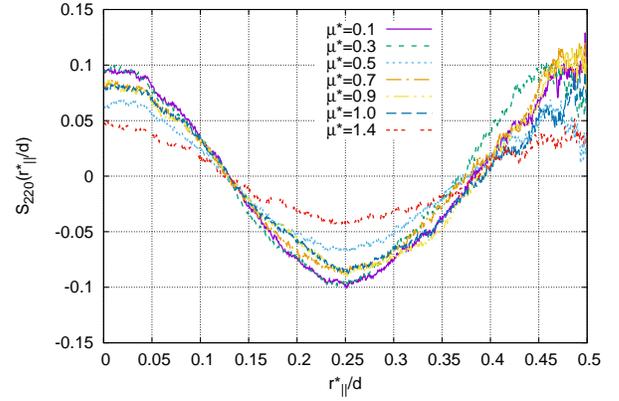}
 \caption{\label{fig:sr220}}
 \end{subfigure}\\
 \begin{subfigure}{0.45\textwidth}
 \includegraphics[width=\textwidth]{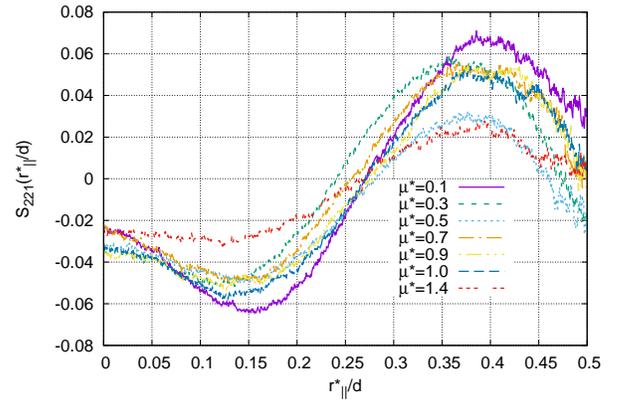}
 \caption{\label{fig:sr221}}
 \end{subfigure}
\caption{Plots of (a) scalar longitudinal orientational correlation functions $S_{220}(r^*_{\parallel}/d)$, (b) pseudo-scalar longitudinal orientational correlation functions $S_{221}(r^*_{\parallel}/d)$.\label{fig:sr}}
\end{center}
\end{figure}
In case of the blue phases obtained, $S_{220}(r^*_{\parallel}/d)$ and $S_{221}(r^*_{\parallel}/d)$ computed separately along each of three mutually perpendicular directions of simulation box, only one is presented in the figure \ref{fig:sr} , show the similar behaviour within statistical errors and indicate that the three orthogonal directions of a cubic blue phase are equivalent to each other. The plot of $S_{220}(r^*_{\parallel}/d)$ ( figure \ref{fig:sr220} ) starts from a positive maximum value where $r^*_{\parallel}/d=0$ and a negative minimum occurs at a separation $r^*_{\parallel}=d/4$, $d$ being twice the sidelength of simulation box along each of the three mutually perpendicular directions of the simulation box. This indicates that the molecules separated by such a distance are more likely perpendicular to each other referring to the existence of double twist cylinders separated by a distance of $1/4^{\text{th}}$ of the lattice constant of cubic unit cell of the blue phase along the chosen reference axis. As $d$ has been taken as equal to the twice of the simulation box it can be said that the simulation box contains only one octant of the whole cubic unit cell of the blue phase. More over, the occurence of another positive maximum at a distance equal to $r^*_{\parallel}=d/2$ refers the presence of two parallelly oriented double twist cylinders at a separation of half the lattice constant of cubic unit cell of the blue phase along the chosen reference axis i.e. molecules separated by such a distance are more likely oriented parallelly. Additionally, in all the blue phases obtained, the plot of $S_{221}(r^*_{\parallel}/d)$ along one of the three mutually perpendicular directions of the simulation box ( figure \ref{fig:sr221} ) has a minimum at a distance $r^*_{\parallel}=d/8$ approximately along the reference axis which indicates that the radius of the double twist cylinders is $1/8^{\text{th}}$ of the lattice constant of the cubic unit cell of the blue phase. All this results are in accordance with results of BP I, as demonstrated by Memmer et al. \cite{memmer2k} , that means all the blue phases obtained in our study are BP I.

\section{Discussions}
Chiral phenomena discovered by Louis Pasteur (1848), has its immense application in pharmaceutical, agricultural, chemical and other technological industries. More than half of the drugs currently in use are chiral compounds. Chirality is a property of an object which is not superimposable on its mirror image. Nearly a century has passed to assess the importance of chirality not only in pharmaceutical and chemical industries but also in biological systems composed of homochiral molecules e.g. DNAs, proteins, cholesterols, carbohydrates, hormones, enzymes etc.

The Blue Phase is an interesting chiral phase which has potential applicational opportunity in the frontiers of condensed matter physics, opto-electronics and biotechnology. The presence of the cubic symmetry in the blue phase liquid crystals makes this phase so interesting and important for experimental research. The blue phase provides the possibility of lasing action in a three dimensional self-assembled photonic liquid crystal \cite{cao}. Nowadays blue phase liquid crystals are the topic of research for their use in developement of next-generation liquid crystal display because of its very small response time and wide field of view. Still there are some difficulties with synthesized blue phase liquid crystals because they are stable for a very short temperature range and their driving voltages are very high. To overcome these difficulties various attempts are being taken such as introduction of polymers and nanoparticles to increase the temperature range \cite{rahman, castles, lee}.

In this present molecular dynamics simulation work we have shown that fine tuning of molecular chirality coupled with dipolar interaction can produce intermolecular forces and torques of suitable amount that can give rise to various chiral phases. The present model not only reproduced cholesteric phase and blue phases but also generated recently discovered chiral smectic phases. Moreover, stabilization of novel chiral bilayered phase has been possible which, we hope, will give new insight to the future experimental realization of this phase for technological application. More importantly the study will enrich our understanding on biomembrane activity, DNA condensed phases and various other liquid crystalline phases found in biological systems.



\end{document}